\documentclass[preprintnumbers,amsmath,amssymb,floatfix,9pt,prd,superscriptaddress
,nofootinbib]{revtex4}
\usepackage{graphicx}
\usepackage{epsfig}
\usepackage{bm}
\usepackage{amsfonts}

\begin{document}

\title{Interacting holographic dark energy with logarithmic correction }

\author{Mubasher Jamil}
\email{mjamil@camp.nust.edu.pk} \affiliation{Center for Advanced
Mathematics and Physics, National University of Sciences and
Technology, Rawalpindi, 46000, Pakistan}

\author{M. Umar Farooq}
\email{m_ufarooq@yahoo.com} \affiliation{Center for Advanced
Mathematics and Physics, National University of Sciences and
Technology, Rawalpindi, 46000, Pakistan}

\begin{abstract}
\textbf{Abstract:} The holographic dark energy (HDE) is considered
to be the most promising candidate of dark energy. Its definition is
originally motivated from the entropy-area relation which depends on
the theory of gravity under consideration. Recently a new definition
of HDE is proposed with the help of quantum corrections to the
entropy-area relation in the setup of loop quantum cosmology. Using
this new definition, we investigate the model of interacting dark
energy and derive its effective equation of state. Finally we
establish a correspondence between generalized Chaplygin gas and
entropy-corrected holographic dark energy.
\end{abstract}
\maketitle


\newpage

\section{Introduction}

Nowadays it is generally accepted in the astrophysics community that
the observable universe is expanding in an accelerated manner due to
the presence of dark energy. The notion of dark energy is strongly
favored by the observations of type Ia supernova, large scale
structure and cosmic microwave background anisotropies
\cite{observation}. From the quantum field theoretic point of view,
the dark energy is nothing but vacuum energy possessing negative
pressure \cite{paddy}. While from the Einstein's general relativity,
the dark energy candidate most probably be the cosmological
constant. However, as is well known, there are two difficulties
arise from the cosmological constant scenario, namely the two famous
cosmological constant problems — the "fine-tuning" problem and the
"cosmic coincidence" problem \cite{pj}. Other alternative candidates
for dark energy \cite{paddy} include the Chaplygin gas, quintessence
and phantom energy  to name a few.

To alleviate the cosmic coincidence and cosmic fine-tuning problems,
the model of dark energy interacting with dark matter has been
proposed \cite{jamil}. Observations suggest that the ratio of energy
densities of matter and dark energy $r_\text{m}$ is closer to unity
at present time, however in pure dark energy models (without
interaction), this ratio must decrease. It leads to another problem
why $r_\text{m}\sim1$ happens at present time? A possible resolution
is to assume that dark energy could decay into matter (or vice
versa) in order to keep a roughly constant ratio in the history of
the universe. With the use of this interaction, the problem of
phantom crossing is also resolved as the Friedmann equations give
stable attractor solution at the present time \cite{zhang}. This
model is also favored by observations of supernova of type Ia where
tight constraints are determined on the coupling parameter of the
dark energy-dark matter interaction \cite{luca}. Besides Einstein's
gravity, the interaction is also modeled in other gravity theories
like the f(R) \cite{popa}, Brans-Dicke \cite{ahmad}, braneworld
\cite{brane}, Horava-Lifshitz \cite{horava} and Gauss-Bonnet
\cite{gauss} gravity.

In recent years, the holographic dark energy has been studied as a
possible candidate for dark energy. It is motivated from the
holographic principle which might lead to the quantum gravity to
explain the events involving high energy scale. In the
thermodynamics of black hole, there is a maximum entropy in a box of
length $L$, commonly termed, the Bekenstein-Hawking entropy bound
$S\sim M_p^2L^2$, which scales as the area of the box $A\sim L^2$
\cite{bek}. To avoid the breakdown of the local quantum field
theory, Cohen et al \cite{cohen} suggested that the entropy for an
effective field theory should satisfy $L^3\Lambda^3\leq S^{3/4}\sim
(M_p L)^{3/2}$. Here $L$ is the size of the region which serves as
the infra-red cut-off while $\Lambda$ is the ultra-violet cut-off.
Incidently this last equation can be re-written in the form
$L^3\rho_\Lambda\leq L M_p^2$, where $\rho_\Lambda\sim\Lambda^4$ is
the energy density corresponding to the zero-point energy and
cut-off $\Lambda$. Thus the total energy in a region of size $L$
cannot exceed the mass of a black hole of the same size. From this
discussion we can deduce that $\rho_\Lambda\leq M_p^2L^{-2}$. This
inequality can be saturated and it becomes $\rho_\Lambda=3n^2
M_p^2L^{-2}$, where $3n^2$ is introduced for convenience. The
holographic dark energy interacting with matter has been widely
discussed in the literature, for instance \cite{setare}. The
resulting interaction also modifies the equations of state of both
dark matter and dark energy. In this paper we determine the
expression for the later quantity and discuss its implications.
Throughout the paper we use units $c=\hbar=1$.

\section{The model}
We assume the background spacetime to be spatially homogeneous and
isotropic given by Friedmann-Robertson-Walker (FRW) metric
\begin{equation}
ds^2=-dt^2+a^2(t)\left[
\frac{dr^2}{1-kr^2}+r^2(d\theta^2+\sin^2\theta d\varphi^2) \right].
\end{equation}
Here $a(t)$ is the dimensionless scale factor which is an arbitrary
function of time and $k$ is the curvature parameter, have dimensions
of $length^{-2}$ and it describes the spatial geometry of spacetime.
For $k=+1,0,-1$, we obtain spatially closed, flat and open FRW
spacetimes respectively. The Einstein field equation representing
the dynamics of FRW spacetime is
\begin{equation}
H^2+\frac{k}{a^2}=\frac{1}{3M_p^2}\left[\rho_\Lambda+\rho_\text{m}\right],
\end{equation}
Here $H\equiv\dot a/a$, is the Hubble constant while $\rho_\Lambda$
and $\rho_\text{m}$ are the energy densities of dark energy and
matter respectively. Also $M_p^2=(8\pi G)^{-1}$, is the modified
Planck mass. One can rewrite Eq. (2) in the dimensionless form as
\begin{equation}
1+\Omega_k=\Omega_\Lambda+\Omega_\text{m},
\end{equation}
where the above density parameters are defined by
\begin{equation}
\Omega_\text{m}=\frac{\rho_\text{m}}{\rho_\text{cr}}=\frac{\rho_\text{m}}{3H^2M_p^2},\
\
\Omega_\Lambda=\frac{\rho_\Lambda}{\rho_\text{cr}}=\frac{\rho_\Lambda}{3H^2M_p^2},\
\ \Omega_k=\frac{k}{(aH)^2}.
\end{equation}
Here $\rho_\text{cr}$ is the critical energy density. The energy
conservation equations for dark energy and matter are
\begin{eqnarray}
\dot{\rho}_\Lambda+3H(\rho_\Lambda+p_\Lambda)&=&-Q,\\
\dot{\rho}_\text{m}+3H\rho_\text{m}&=&Q,
\end{eqnarray}
where $Q$ is an interaction term which can be an arbitrary function
of cosmological parameters like the Hubble parameter and energy
densities $Q(H\rho_\Lambda,H\rho_\text{m})$. The simplest choice is
$Q\simeq H(\rho_\Lambda+\rho_\text{m})$ up to the linear order in
energy densities. The relation for the interaction term can be
saturated by introducing a coupling parameter $b^2$ by $Q=
3b^2H(\rho_\Lambda+\rho_\text{m})$ \cite{zhang}. Observations of
cosmic microwave background and galactic clusters show that the
coupling parameter $b^2<0.025$, i.e. a small but positive constant
of order unity \cite{luca}, a negative coupling parameter is avoided
due to violation of thermodynamical laws. It should be noted that
the ideal interaction term must be motivated from the theory of
quantum gravity. In the absence of such a theory, we rely on pure
dimensional basis for choosing an interaction $Q$. In literature,
various forms of $Q$ are proposed \cite{chimento}. The effective
equations of state for dark energy and matter are defined by
\cite{ali}
\begin{equation}
\omega_\Lambda^\text{eff}=\omega_\Lambda+\frac{\Gamma}{3H},\ \
\omega_\text{m}^\text{eff}=-\frac{1}{r_\text{m}}\frac{\Gamma}{3H}.
\end{equation}
Here $r_\text{m}=\rho_\text{m}/\rho_\Lambda$, and
$\Gamma=Q/\rho_\Lambda=3H(1+r_\text{m})$, is the decay rate of dark
energy into matter. Making use of (7) in (5) and (6), we have
\begin{eqnarray}
\dot{\rho}_\Lambda+3H(1+\omega_\Lambda^\text{eff})\rho_\Lambda&=&0,\\
\dot{\rho}_\text{m}+3H(1+\omega_\text{m}^\text{eff})\rho_\text{m}&=&0.
\end{eqnarray}

Notice that the definition and derivation of holographic dark energy
($\rho_\Lambda=3n^2 M_p^2L^{-2}$) depends on the entropy-area
relationship $S\sim A\sim L^2$ in Einstein's gravity. Here $A$
represents the area of the horizon. However, this definition can be
`corrected' from the inclusion of quantum effects, motivated from
the loop quantum gravity (LQG). The quantum corrections provided to
the entropy-area relationship leads to the curvature correction in
the Einstein-Hilbert action and vice versa \cite{zhu}. The corrected
entropy is $S=(A/4G) + \tilde{\gamma}\ln (A/4G)+\tilde\beta$, where
$\tilde{\gamma}$ and $\tilde\beta$ are constants of order unity. The
exact values of these constants are not yet determined and still an
open issue in loop quantum cosmology. These corrections arise in the
black hole entropy in LQG due to thermal equilibrium fluctuations
and quantum fluctuations \cite{carlo}. The entropy corrected
holographic dark energy (ECHDE) is given by \cite{hao}
\begin{equation}
\rho _{\Lambda }=3n^2M_{p}^{2}L^{-2}+\gamma L^{-4}\ln
(M_{p}^{2}L^{2})+\beta L^{-4}.
\end{equation}
Here $n^2$, $\gamma$ and $\beta$ are dimensionless constants of
order unity. Note that choosing $\gamma=\beta=0$, yields the
well-known holographic dark energy. There is a possibility that the
constants involved in (10) could be time dependent. In a recent
study, Xu \cite{xu} has considered $n^2=n^2(t)$ and hence deduced
that the resulting equation of state of holographic dark energy is
consistent with the observations of SN Ia and BAO. Hence one can
infer the same time dependence for $\gamma$ and $\beta$ as well,
however, we shall treat all the parameters $n^2$, $\gamma$ and
$\beta$ to be constants. If we choose $L$ as the size of the
universe, for instance the Hubble horizon $H^{-1}$, the resulting
$\rho _{\Lambda }$ is comparable to the observational density of
dark energy. However, Hsu \cite{hsu} pointed out that in this case
the resulting equation-of-state parameter (EoS) is equal to zero,
which cannot accelerate the expansion of our universe. One could
also employ the particle horizon as a desirable cut-off but it
turned out that it yielded $w_\Lambda>-1/3$, a form of exotic matter
that could not derive accelerated expansion. To get an accelerating
universe, Li \cite{li} proposed that $L$ should be the future event
horizon $R_h$. Li defined it as
\begin{equation}
L=a(t)\frac{\text{sinn}(\sqrt{|k|}y)}{\sqrt{|k|}},\ \ \
y=\frac{R_h}{a(t)},
\end{equation}
where $R_h$ is the size of the future event horizon defined as
\begin{equation}
R_h=a(t)\int\limits_t^\infty\frac{dt^\prime}{a(t^\prime)}=a(t)\int\limits_0
^{r_1}\frac{dr}{\sqrt{1-kr^2}}.
\end{equation}
The last integral has the explicit form as
\begin{equation}\int\limits_0^{r_1}\frac{dr}{\sqrt{1-kr^2}}=\frac{1}{\sqrt{|k|}}
\text{sinn}^{-1}(\sqrt{|k|}r_1)=
\begin{cases} \text{sin}^{-1}(r_1) , & \, \,k=+1,\\
             r_1, & \, \,  k=0,\\
             \text{sinh}^{-1}(r_1), & \, \,k=-1.\\
\end{cases}\end{equation}

In recent years, some new infrared cut-offs have been proposed in
the literature. In \cite{granda}, the authors have added the square
of the Hubble parameter and its time derivative within the
definition of holographic dark energy. While in \cite{sad}, the
authors propose a linear combination of particle horizon and the
future event horizon. However, in this paper we stick to Li's
proposal.

Using the definitions of $\Omega_\Lambda$ and $\rho_\text{cr}$, we
obtain a useful relation
\begin{equation}
HL=\sqrt{\frac{3n^2M_p^2+\gamma L^{-2}\ln(M_p^2L^2)+\beta
L^{-2}}{3M_p^2\Omega_\Lambda}}.
\end{equation}
Differentiating $L$ with respect to time $t$ and using (14) yields
\begin{equation}
\dot L=\sqrt{\frac{3n^2M_p^2+\gamma L^{-2}\ln(M_p^2L^2)+\beta
L^{-2}}{3M_p^2\Omega_\Lambda}}-\text{cosn}(\sqrt{|k|}y),
\end{equation}
where
\begin{equation}\text{cosn}(\sqrt{|k|}y)=
\begin{cases} \cos y  & \, \,k=+1,\\
             1 & \, \,  k=0,\\
             \cosh y & \, \,k=-1.\\
\end{cases}\end{equation}
 Differentiating (10) with respect to $t$ gives
\begin{eqnarray}
\dot{\rho} _{\Lambda }&=&\Big[2\gamma L^{-5}-4\gamma L^{-5}\ln
(M_{p}^{2}L^{2})-4\beta
L^{-5}-6n^2M_{p}^{2}L^{-3}\Big]\Big[\sqrt{\frac{3n^2M_p^2+\gamma
L^{-2}\ln(M_p^2L^2)+\beta
L^{-2}}{3M_p^2\Omega_\Lambda}}-\text{cosn}(\sqrt{|k|}y)\Big].\nonumber\\
\end{eqnarray}
Making use of (17) in (5) gives
\begin{eqnarray}
w_{\Lambda } &=&-1-\frac{ 2\gamma L^{-2} -4\gamma L^{-2}\ln
(M_{p}^{2}L^{2})-4\beta L^{-2}-6n^2M_{p}^{2}}{3(3n^2M_{p}^{2}+\gamma
L^{-2}\ln (M_{p}^{2}L^{2})+\beta
L^{-2})}\Big[1-\sqrt{\frac{3M_p^2\Omega_\Lambda}{3n^2M_p^2+\gamma
L^{-2}\ln(M_p^2L^2)+\beta
L^{-2}}}\text{cosn}(\sqrt{|k|}y)\Big]\nonumber\\&&-\frac{b^{2}(1+\Omega
_k)}{\Omega _{\Lambda}}.
\end{eqnarray}
Using (18) in (7) gives
\begin{eqnarray}
w_{\Lambda }^\text{eff}&=&-1-\frac{ 2\gamma L^{-2} -4\gamma
L^{-2}\ln (M_{p}^{2}L^{2})-4\beta
L^{-2}-6n^2M_{p}^{2}}{3(3n^2M_{p}^{2}+\gamma L^{-2}\ln
(M_{p}^{2}L^{2})+\beta
L^{-2})}\Big[1-\sqrt{\frac{3M_p^2\Omega_\Lambda}{3n^2M_p^2+\gamma
L^{-2}\ln(M_p^2L^2)+\beta
L^{-2}}}\text{cosn}(\sqrt{|k|}y)\Big].\nonumber\\
\end{eqnarray}
The above expression represents the effective equation of state for
the entropy corrected holographic dark energy interacting with
matter. An interesting case arises when the FRW universe is
spatially flat $k=0$, then Eq. (19) gives $w_{\Lambda
}^\text{eff}=-1$. However in the non-flat case, the complete
expression on right hand side in (19) will play crucial role and
phantom crossing will be possible for selected choices of
parameters. Another important implication of ECHDE besides phantom
energy is the cosmological inflation in the early universe: in this
case the Hubble horizon $H^{-1}$ and the future event horizon $R_h$
will coincide i.e. $L=R_h=H^{-1}$ (collectively implying $H=$
constant during the inflation era). Therefore the equation of state
of ECHDE during the inflation era will be
\begin{eqnarray}
w_{\Lambda }^\text{eff}&=&-1-\frac{ 2\gamma H^{2} -4\gamma H^{2}\ln
(M_{p}^{2}H^{-2})-4\beta H^{2}-6n^2M_{p}^{2}}{3(3n^2M_{p}^{2}+\gamma
H^{2}\ln (M_{p}^{2}H^{-2})+\beta
H^{2})}\Big[1-\sqrt{\frac{3M_p^2\Omega_\Lambda}{3n^2M_p^2+\gamma
H^{2}\ln(M_p^2H^{-2})+\beta
H^{2}}}\text{cosn}(\sqrt{|k|}y)\Big].\nonumber\\
\end{eqnarray}
After the end of the inflationary phase, the universe subsequently
enters in the radiation and then matter dominated eras. Since during
the later two stages, the dark energy is not the dominant component
of the total cosmic energy density, one can safely take
$\gamma=\beta=0$, i.e. the correction terms can work only in the
inflationary or in the late time acceleration phases. For the later
case, the infra-red cut-off will be $L=R_h$ ($R_h\neq H^{-1}$ and
$R_h\neq$ constant). Notice that if the accelerated expansion is due
to phantom energy then the future event horizon is a decreasing
function of time ($\dot {R}_h\leq0$) while for all other cases,
$\dot{R}_h>0$ \cite{sadjadi}.

\section{Correspondence between generalized Chaplygin gas and
ECHDE}

Since there are numerous candidates for dark energy, it is essential
to know how these various candidates are related to each other. In
this connection, we proceed to obtain a correspondence between
generalized Chaplygin gas (GCG) and ECHDE. One of the possible
candidates for dark energy is the GCG which is the generalization of
the Chaplygin gas \cite{kamen}. It has an amazing property of
interpolating the evolution of the universe from the dust phase to
the accelerated phase of the universe and hence best fits the
observational data \cite{lu}. The model of GCG as well as its
further generalization have been extensively studied in the
literature \cite{general}. The GCG is defined as \cite{bento}
\begin{equation}
p_\Lambda=-\frac{D}{\rho_\Lambda^\alpha}.
\end{equation}
Here $D$ is a positive constant and $\alpha$ is also a constant.
Notice that fixing $\alpha=1$ yields the Chaplygin gas. It is shown
in \cite{gorini} that the matter power spectrum is compatible with
the observed one only for $\alpha < 10^{-5}$, which makes the
generalized Chaplygin gas practically indistinguishable from the
standard cosmological model with cosmological constant
($\Lambda$CDM). In \cite{zhang1}, the Chaplygin inflation has been
investigated in the context of loop quantum cosmology and it is
shown that the parameters of the Chaplygin inflation model are
consistent with the WMAP 5-year results.

The density evolution of GCG is given by
\begin{equation}
\rho_\Lambda=\Big[D+\frac{B}{a^{3(1+\alpha)}}
\Big]^{\frac{1}{1+\alpha}},
\end{equation}
where $B$ is a constant of integration. We proceed with the
reconstruction of the potential and the dynamics of the scalar field
in the light of the ECHDE. The energy density and the pressure of
the homogeneous and time dependent scalar field $\Phi$ are given by
\begin{eqnarray}
\rho_\Lambda&=&\frac{\sigma}{2}\dot\Phi^2+V(\Phi),\\
p_\Lambda&=&\frac{\sigma}{2}\dot\Phi^2-V(\Phi).
\end{eqnarray}
Here $\sigma=-1$ corresponds to the phantom while $\sigma=+1$
represents the standard scalar field which represent the
quintessence field, also $V(\Phi)$ is the potential. In this case
$w_{\Lambda }$ is given by
\begin{equation}
w_\Lambda=\frac{p_\Lambda}{\rho_\Lambda}=\frac{\sigma\dot\Phi^2-2V(\Phi)}
{\sigma\dot\Phi^2+2V(\Phi)}.
\end{equation}
We observe that it results in the violation of the null energy
condition $\rho_\Lambda+p_\Lambda=\sigma\dot{\Phi}^2>0$, if
$\sigma=-1$. Since the null energy condition is the basic condition,
its violation yields other standard energy conditions to be violated
likewise dominant energy condition ($\rho_\Lambda>0$,
$\rho_\Lambda\geq|p_\Lambda|$) and the strong energy condition ($
\rho_\Lambda+p_\Lambda>0 $, $ \rho_\Lambda+3p_\Lambda>0 $). Due to
the energy condition violations, it makes the failure of cosmic
censorship conjecture and theorems related to black hole
thermodynamics. The prime motivation to introduce this weird concept
in cosmology does not come from the theory but from the
observational data. According to the forms of dark energy density
and pressure (23) and (24), one can easily obtain the kinetic energy
and the scalar potential terms as
\begin{eqnarray}
\dot{\Phi}^2&=&\frac{1}{\sigma}(1+\omega_\Lambda)\rho_\Lambda,\\
V(\Phi)&=&\frac{1}{2}(1-\omega_\Lambda)\rho_\Lambda.
\end{eqnarray}
But we know that
\begin{equation}
w_{\Lambda }=-\frac{D}{\rho_\Lambda ^{\alpha
+1}}=-\frac{D}{(3n^2M_{p}^{2}L^{-2}+\gamma L^{-4}\ln
(M_{p}^{2}L^{2})+\beta L^{-4}) ^{\alpha +1}}.
\end{equation}
Therefore using (18) and (28) yields the value of the parameter
\begin{eqnarray}
D &=&(3n^2M_{p}^{2}L^{-2}+\gamma L^{-4}\ln (M_{p}^{2}L^{2})+\beta
L^{-4})^{\alpha +1}\Big[1+\frac{ 2\gamma L^{-2} -4\gamma L^{-2}\ln
(M_{p}^{2}L^{2})-4\beta L^{-2}-6n^2M_{p}^{2}}{3(3n^2M_{p}^{2}+\gamma
L^{-2}\ln (M_{p}^{2}L^{2})+\beta
L^{-2})}\nonumber\\&&\times\Big\{1-\sqrt{\frac{3M_p^2\Omega_\Lambda}{3n^2M_p^2+\gamma
L^{-2}\ln(M_p^2L^2)+\beta
L^{-2}}}\text{cosn}(\sqrt{|k|}y)\Big\}+\frac{b^{2}(1+\Omega
_k)}{\Omega _{\Lambda}}\Big].
\end{eqnarray}
Eq. (22) implies $B=a^{3(\alpha +1)}(\rho _{\Lambda }^{\alpha
+1}-D)$, which after using (29) takes the form
\begin{eqnarray}
B &=&-[a^{3}(3n^2M_{p}^{2}L^{-2}+\gamma L^{-4}\ln
(M_{p}^{2}L^{2})+\beta L^{-4})]^{\alpha +1}\Big[ \frac{ 2\gamma
L^{-2} -4\gamma L^{-2}\ln (M_{p}^{2}L^{2})-4\beta
L^{-2}-6n^2M_{p}^{2}}{3(3n^2M_{p}^{2}+\gamma L^{-2}\ln
(M_{p}^{2}L^{2})+\beta
L^{-2})}\nonumber\\&&\times\Big\{1-\sqrt{\frac{3M_p^2\Omega_\Lambda}{3n^2M_p^2+\gamma
L^{-2}\ln(M_p^2L^2)+\beta
L^{-2}}}\text{cosn}(\sqrt{|k|}y)\Big\}+\frac{b^{2}(1+\Omega
_k)}{\Omega _{\Lambda}}  \Big].
\end{eqnarray}%
Using Eqs. (26),(27),(29) and (30), we obtain the kinetic and
potential terms
\begin{eqnarray}
\sigma\dot\Phi^2 &=&-(3n^2M_{p}^{2}L^{-2}+\gamma L^{-4}\ln
(M_{p}^{2}L^{2})+\beta L^{-4})\Big[\frac{ 2\gamma L^{-2} -4\gamma
L^{-2}\ln (M_{p}^{2}L^{2})-4\beta
L^{-2}-6n^2M_{p}^{2}}{3(3n^2M_{p}^{2}+\gamma L^{-2}\ln
(M_{p}^{2}L^{2})+\beta
L^{-2})}\nonumber\\&&\times\Big\{1-\sqrt{\frac{3M_p^2\Omega_\Lambda}{3n^2M_p^2+\gamma
L^{-2}\ln(M_p^2L^2)+\beta
L^{-2}}}\text{cosn}(\sqrt{|k|}y)\Big\}+\frac{b^{2}(1+\Omega
_k)}{\Omega _{\Lambda}}\Big],\\
2V(\Phi ) &=&(3n^2M_{p}^{2}L^{-2}+\gamma L^{-4}\ln
(M_{p}^{2}L^{2})+\beta L^{-4})\Big[2+\frac{ 2\gamma L^{-2} -4\gamma
L^{-2}\ln (M_{p}^{2}L^{2})-4\beta
L^{-2}-6n^2M_{p}^{2}}{3(3n^2M_{p}^{2}+\gamma L^{-2}\ln
(M_{p}^{2}L^{2})+\beta
L^{-2})}\nonumber\\&&\times\Big\{1-\sqrt{\frac{3M_p^2\Omega_\Lambda}{3n^2M_p^2+\gamma
L^{-2}\ln(M_p^2L^2)+\beta
L^{-2}}}\text{cosn}(\sqrt{|k|}y)\Big\}+\frac{b^{2}(1+\Omega
_k)}{\Omega _{\Lambda}}\Big].
\end{eqnarray}
We can write $\dot\Phi=\Phi'H,$ where prime denotes differentiation
with respect to $\ln a$. Hence from (31) we can write
\begin{eqnarray}
\Phi(a)-\Phi(a_0)&=&\int\limits_0^{\ln
a}\frac{1}{H}\Big[-\frac{1}{\sigma}(3n^2M_{p}^{2}L^{-2}+\gamma
L^{-4}\ln (M_{p}^{2}L^{2})+\beta
L^{-4})\nonumber\\&\;&\times\Big(\frac{ 2\gamma L^{-2} -4\gamma
L^{-2}\ln (M_{p}^{2}L^{2})-4\beta
L^{-2}-6n^2M_{p}^{2}}{3(3n^2M_{p}^{2}+\gamma L^{-2}\ln
(M_{p}^{2}L^{2})+\beta
L^{-2})}\nonumber\\&&\times\Big\{1-\sqrt{\frac{3M_p^2\Omega_\Lambda}{3n^2M_p^2+\gamma
L^{-2}\ln(M_p^2L^2)+\beta
L^{-2}}}\text{cosn}(\sqrt{|k|}y)\Big\}+\frac{b^{2}(1+\Omega
_k)}{\Omega _{\Lambda}}\Big)\Big]^{1/2}d\ln a.\nonumber\\
\end{eqnarray}
Here $a_0$ denotes the present value of the scale factor. Also
expression (32) represents the reconstructed potential. We will also
comment that the  present analysis can be performed for entropy
corrected new agegraphic dark enrgy (ECNADE) \cite{hao} whose
definition closely mimics to that of ECHDE (10). The former is
defined by
\begin{equation}
\rho_\Lambda=3n^2M_p^2\eta^2+\beta\eta^{-4}.
\end{equation}
By comparing the above definition (34) with (10), we note that
$\gamma=0$ and $L$ is replaced with the conformal time
$\eta\equiv\int\limits_0^a\frac{d\tilde{a}}{H\tilde{a}^2}$. However
there are several disadvantages with the agegraphic dark energy: it
can not generate the inflation era in the early universe unlike the
holographic dark energy; it cannot produce a phantom dominated
universe since its equation of state parameter is always greater
than $-1$; its energy density decreases with time unlike any other
dark energy candidate; quantum corrections are generally ignorable
and it worse fits with the observational data \cite{hao,lu}.
Moreover in this paper we have restricted our analysis for the
entropy corrections up to second order although these corrections
can be extended to higher orders and the present analysis can be
generalized to the desired order of correction.

\section{Conclusion}
In this paper we have investigated the model of interacting dark
energy with the inclusion of entropy corrections to the holographic
dark energy. These corrections are motivated from the LQG which is
one of the promising theories of quantum gravity. Among various
candidates to play the role of the dark energy, the generalized
Chaplygin gas has emerged as a possible unification of dark matter
and dark energy. This unification arises since its cosmological
evolution is similar to an initial dust like matter and a
cosmological constant for late times. In this paper, by considering
an interaction between the entropy corrected holographic dark energy
and matter, we have obtained the equation of state for the
interacting entropy corrected holographic dark energy energy density
in the non-flat universe. We have considered a correspondence
between the holographic dark energy density and interacting
generalized Chaplygin gas energy density in FRW universe. Finally we
have reconstructed the potential of the scalar field which describe
the generalized Chaplygin cosmology.

\end{document}